\documentclass[aps,pre,amsmath,amssymb,reprint,superscriptaddress,longbibliography]{revtex4-2}

\usepackage{graphicx}
\usepackage{bm}
\usepackage{xcolor}
\usepackage[colorlinks=true,linkcolor=blue,citecolor=blue,urlcolor=blue]{hyperref}
\usepackage{verbatim}
\usepackage{braket}

\usepackage{float}
\usepackage{dcolumn}

\usepackage[english]{babel}
\usepackage{xfrac}

\newcommand{\Op}{\hat{S}}
\newcommand{\spps}{$sp\,p's'$}

\begin{document}

\title{A random sampling method to test the weak Eigenstate Thermalization Hypothesis}

\author{Nicolaus A.\ Chlanda}
\affiliation{Department of Physics, Bryn Mawr College,
             Bryn Mawr, Pennsylvania 19010, USA}
\author{John Keim}
\author{Isabel Martinez-Robles}
\affiliation{Department of Physics and Astronomy, Ursinus College,
             Collegeville, Pennsylvania 19426, USA}

\author{Sarah E. Spielman}
\affiliation{Department of Physics, Bryn Mawr College,
             Bryn Mawr, Pennsylvania 19010, USA}

\author{Thomas J.\ Carroll}
\affiliation{Department of Physics and Astronomy, Ursinus College,
             Collegeville, Pennsylvania 19426, USA}
\author{Michael W.\ Noel}             
\affiliation{Department of Physics, Bryn Mawr College,
             Bryn Mawr, Pennsylvania 19010, USA}

\date{\today}

\begin{abstract}
We introduce a random sampling method for testing the weak Eigenstate Thermalization Hypothesis (wETH) that is particularly useful for large, sparse Hamiltonians. We test the method on a well-studied Rydberg atom system that exhibits both two- and three-body dipole-dipole interactions. Our results show that our method allows for the study of system sizes significantly beyond the limits of exact diagonalization. We find that the fluctuations in eigenstate expectation values decay as the Hilbert space dimension $D^{-1/2}$, consistent with wETH and in agreement with previous numerical results.
\end{abstract}

\maketitle

\section{Introduction}

Understanding how isolated quantum systems evolve to thermal equilibrium is of central importance to the project of deriving statistical physics from fundamental quantum laws. The eigenstate thermalization hypothesis (ETH) has emerged as the leading explanation~\cite{deutsch_QuantumStatisticalMechanics_1991,srednicki_chaos_1994,rigol_ThermalizationItsMechanism_2008}. The ETH posits that within a narrow energy shell, all energy eigenstates are already thermal. That is, energy eigenstates with similar eigenvalues will have similar expectation values. 

ETH has been numerically tested, showing that it holds in a wide variety of contexts~\cite{kim_TestingWhetherAll_2014,sugimoto_TestEigenstateThermalization_2021,sugimoto_EigenstateThermalizationLongRange_2022,steinigeweg_PushingLimitsEigenstate_2014,beugeling_FinitesizeScalingEigenstate_2014,ikeda_HowAccuratelyCan_2015}. Typically, such experiments must solve for all of the eigenstates to test the validity of ETH. However, violations of ETH have also been discovered, including many-body localization~\cite{nandkishore_ManyBodyLocalizationThermalization_2015,abanin_ColloquiumManybodyLocalization_2019,luschen_ObservationSlowDynamics_2017}, quantum many-body scar states~\cite{turner_QuantumScarredEigenstates_2018,turner_WeakErgodicityBreaking_2018,spielman_QuantumManybodyScars_2024}, and Hilbert space fragmentation~\cite{sala_ErgodicityBreakingArising_2020}. This has led to the formulation of the weak ETH (wETH), in which only the vast majority of the energy eigenstates are required to be thermal.

Our work is motivated by Hamiltonians that are known to violate ETH due to the presence of quantum many-body scar states. Scar states are rare eigenstates with atypical expectation values and abnormally low entanglement entropy. Thermalization is not guaranteed in such cases, as the initial state may have significant overlap with the scar states. However, if these Hamiltonians satisfy wETH, we can still make predictions about the thermal equilibrium state~\cite{reimann_DynamicalTypicalityApproach_2018}.

The wETH requires that the fraction of atypical eigenstates should become vanishingly small in the thermodynamic limit. Thus, we do not need to check every eigenstate. Rather, we develop a method to statistically sample the eigenstates to test whether a Hamiltonian satisfies wETH. To do so, we calculate the variance, among the sampled eigenstates, of an experimentally relevant expectation value. Our method is particularly useful when the Hamiltonian is sparse and too large to diagonalize exactly.

We apply our method to a recently studied system of ultracold Rydberg atoms, which exchange energy via two- and three-body dipole-dipole interactions. In the case of three-body interactions, experiment suggests that the system fails to thermalize~\cite{liu_TimeDependenceFewBody_2020} and numerical results indicate that quantum many-body scar states are to blame~\cite{spielman_QuantumManybodyScars_2024}. For system sizes that are not too large, we compare our new method to exact diagonalization. Our sampling method is particularly important for the three-body interaction case, as exact diagonalization cannot probe large enough system sizes to make a definitive conclusion about the status of wETH.

We find that these Hamiltonians indeed still obey wETH, so that techniques like quantum dynamical typicality~\cite{reimann_TypicalityGeneralizedMicrocanonical_2007, bartsch_DynamicalTypicalityQuantum_2009,steinigeweg_PushingLimitsEigenstate_2014,heitmann_SelectedApplicationsTypicality_2020, reimann_DynamicalTypicalityIsolated_2018,mitric_DynamicalQuantumTypicality_2025} can be used to predict the thermal state, as was done in~\cite{spielman_ThermalizationStarkManifold_2026}. We also find that the variance scales as the Hamiltonian dimension $D^{-1/2}$, in agreement with previous  studies~\cite{beugeling_FinitesizeScalingEigenstate_2014,steinigeweg_PushingLimitsEigenstate_2014,ikeda_HowAccuratelyCan_2015,dooley_ExtremeManybodyScarring_2022}.

The remainder of this work is organized as follows. In Sec.~\ref{sec:method}, we present our metric for testing wETH and the details of our algorithm.  Additional details and discussion of our implementation choices are provided in the appendix. We discuss the two- and three-body dipole-dipole interacting Hamiltonians that we test with our method in Sec.~\ref{sec:model}. We report our main results on the validity of wETH in our system in Sec.~\ref{sec:results}. Finally, we conclude with a discussion of our results and future prospects for our method in Sec.~\ref{sec:conclusion}.

\section{Method}\label{sec:method}

Our method starts by using the kernel polynomial method (KPM) to estimate the density of states $\rho(E)$~\cite{weisse_KernelPolynomialMethod_2006}. Armed with $\rho(E)$, we randomly select an energy within the middle of the spectrum. We then use Chebyshev filter diagonalization (CFD) to find nearby eigenstates and eigenvalues. Finally, to test wETH, we compute the variance of an expectation value and check how quickly it decreases with system size.

\subsection{wETH Indicator}\label{sec:wETH-indicator}
According to the ETH, eigenstates with similar energies will have similar expectation values. In other words, the variance of these expectation should vanish as the system size increases. To evaluate the scaling of the variance with system size, we choose an operator $\hat{S}$, and define the variance as
\begin{equation}
\sigma_{\Delta S}^2 = \frac{1}{N_K}\sum_{E_i \in K}\bigl(S_i - \langle S\rangle\bigr)^{2},
\label{eq:Delta_def}
\end{equation}
where $N_K$ is the number of sampled eigenstates in the energy window $K=[E_0-\Delta E, E_0+\Delta E]$, $E_i$ is the energy of the eigenstate $\ket{E_i}$, $\langle S\rangle$ is the average expectation value of all sampled states, and $S_i = \langle E_i | \Op | E_i\rangle$ is the expectation value for the $i^{th}$ sampled state. For our energy, window we use the middle 5\% of the spectrum for all system sizes studied.

The variance $\sigma_{\Delta S}$ of Eq.~(\ref{eq:Delta_def}) has been used in multiple numerical tests of ETH~\cite{beugeling_FinitesizeScalingEigenstate_2014}. Collectively, these studies support the notion of ETH and find that the variance scales as the Hamiltonian dimension $D^{-1/2}$. However, this result disagrees with Huang~\cite{huang_FinitesizeScalingAnalysis_2022}, who analytically derived that the variance should scale as a power law in the system size for generic Hamiltonians.

\subsection{Sampling Algorithm}\label{sec:algorithm}
The first step of the sampling algorithm is to use the KPM, which constructs a smooth approximation of the density of states $\rho(E)$ from stochastic Chebyshev moments. We then normalize $\rho(E)$ and draw target energies from the middle 5\% of the spectrum by rejection sampling. A candidate energy $z_\ell$ is proposed uniformly across the interior region, a number $u$ is drawn uniformly from $[0,1]$, and the candidate is accepted if $u<\rho(z_\ell)/\rho_{\max}$, where $\rho_{\max}$ is the maximum of the reconstructed density over the region. Each accepted $z_\ell$ becomes the center of one CFD solve. 

Around each accepted energy $z_\ell$ we open a small adaptive bracket $[E_{\rm lo},E_{\rm hi}]\ni z_\ell$ chosen to contain approximately $N_T$  eigenvalues estimated from our KPM spectrum. CFD constructs a polynomial $p(\widetilde{H})$ that acts as a near-projector onto the bracket, large for eigenvalues inside $[E_{\rm lo},E_{\rm hi}]$ and strongly damped outside, so that applying it to a random vector amplifies the wanted spectral components and suppresses the rest. We start from $N_S$ random gaussian search vectors, filter each through $p(\widetilde{H})$ to project it into the bracket subspace, and orthonormalize the filtered set with SVQB~\cite{stathopoulos_BlockOrthogonalizationProcedure_2002}. Projecting $\widetilde{H}$ onto the resulting orthonormal basis yields a small dense matrix of dimension $N_S$ whose eigenpairs, the Rayleigh--Ritz pairs, approximate the eigenpairs of $\widetilde{H}$ inside the bracket. After enough iterations, where $\|\widetilde{H}v_j-\theta_j v_j\|_2/\|v_j\|_2\le \tau_{\rm res}$, where $\tau_{\rm res}$ is a tolerance, $v_j$ is the eigenvector and $\theta_j$ is the eigenvalue, the expectation values $X_i$ of the retained Ritz pairs are computed. After enough expectation values have been collected, we compute the standard deviation. A flowchart of the method can be found at Fig. \ref{fig:method}. More details on the implementation choices can be found in appendix~\ref{app:impl}.

Ideally, for true random sampling, we would obtain the single eigenpair closest to each sampled energy. However, CFD calculates the eigenpairs in some range and, thus, it is significantly more computationally efficient to obtain $\sim N_T$ eigenpairs at each sampled energy. This is only problematic if eigenstates near to each other in energy are correlated. Based on the success of random matrix theory, the coefficients of mid-spectrum eigenstates have generally considered to be effectively random; however, recent work has uncovered subtle correlations~\cite{haque_EntanglementMidspectrumEigenstates_2022,kliczkowski_AverageEntanglementEntropy_2023,rodriguez-nieva_QuantifyingQuantumChaos_2024,hahn_EigenstateCorrelationsEigenstate_2024,langlett_EntanglementPatternsQuantum_2025}. The smallness of these correlations, along with the agreement of our exact and statistical results in Sec.~\ref{sec:results}, give us some confidence that correlations among the eigenstates are not an issue here.

\begin{figure}[!htbp]
\centering
\includegraphics[width=0.6\columnwidth]{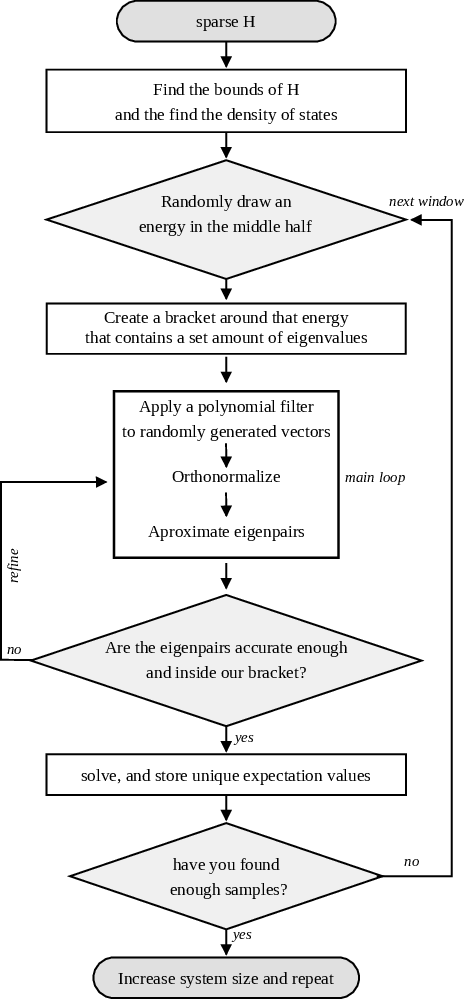}
\caption{A flowchart of the general method. For a sparse Hamiltonian $H$ find the normalized density of states $\rho(E)$ and begin sampling interior energies. Around each accepted energy an adaptive bracket is created, in which we then create a set of random gaussian vectors. We now begin the main loop: apply a polynomial to vectors, orthonormalize, and solve for  eigenpairs. If they are accurate enough, and inside our bracket we have created, their expectation values are stored. Once we have hit our desired number, we first compute the standard deviation of expectation values and continue to the next system size if available. Once we have tested enough systems, we then plot the standard deviation as a function of system size.}
\label{fig:method}
\end{figure}

\section{Model Rydberg system}\label{sec:model}
\begin{figure}
\centering
\includegraphics{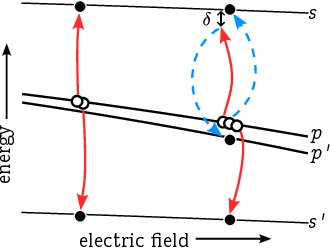}
\caption{Energy level diagram.
of the \spps{} Rydberg system showing the two- and three-body interactions. The horizontal axis is the applied electric field while the vertical axis is energy. Pairs of solid red or dashed blue arrows represent dipole-dipole interactions. Open circles indicate initial states and closed circles indicate final states. The detuning $\delta$ of the three-body interaction is labeled by a double-arrow at the top-center.}
\label{fig:spps}
\end{figure}

For principle quantum numbers $32 < n < 38$, rubidium atoms in $np$ energy levels can resonantly exchange energy via two- and three-body dipole-dipole interactions~\cite{faoro_BorromeanThreebodyFRET_2015,tretyakov_ObservationBorromeanThreeBody_2017,liu_TimeDependenceFewBody_2020}. These interactions are tuned into resonance by the Stark effect, as shown in Fig.~\ref{fig:spps}, where we also define the labels for our states in this \spps{} model. At the two-body resonance on the left of Fig.~\ref{fig:spps}, where $E_s-E_p=E_p-E_{s'}$, we have
\begin{equation}
    p + p \leftrightarrow s + s'.
\end{equation}
If the applied electric field is increased, the two-body interaction can be tuned out of resonance by the amount $\delta=E_p-E_{p'}$. This defect can be accounted for by recruiting a third atom to interact via $s+p\leftrightarrow p'+s$, which is detuned by $-\delta$, yielding the resonant three-body interaction
\begin{equation}
    p + p + p \leftrightarrow s + s' +p'.
\end{equation}
At both the two-body and three-body resonant fields, there are also the hopping interactions
\begin{eqnarray}
    p + s &\leftrightarrow& s + p\label{eq:spps-ar1}\\
    p + s' &\leftrightarrow& s' + p\\
    p' + s &\leftrightarrow& s + p'\\
    p' + s' &\leftrightarrow& s' + p'.\label{eq:spps-ar}
\end{eqnarray}

For $n \lesssim  34$, the two- and three-body resonant fields are separated enough that we can treat the two cases independently. The two-body Hamiltonian operator is
\begin{equation}
\begin{split}
\hat{H}_2 ={}& \sum_{i\neq j}\!\left(\hat{\sigma}^{i}_{ps}\,\hat{\sigma}^{j}_{ps'} + \mathrm{H.c.}\right)\!\frac{\mu\nu}{R_{ij}^3} \\
&+ \sum_{i\neq j}\!\left(\mu^2\,\hat{\sigma}^{i}_{ps}\hat{\sigma}^{j}_{sp} + \nu^2\,\hat{\sigma}^{i}_{ps'}\hat{\sigma}^{j}_{s'p}\right)\!\frac{1}{R_{ij}^3}.
\end{split}
\label{eq:H2}
\end{equation}
The three-body Hamiltonian operator is
\begin{equation}
\begin{split}
\hat{H}_3 ={}& \sum_{i\neq j\neq k}\!\left(\hat{\sigma}^{i}_{ps}\,\hat{\sigma}^{j}_{ps'}\,\hat{\sigma}^{k}_{pp'} + \mathrm{H.c.}\right)\!V_3^{ijk} \\
&+ \sum_{i\neq j}\Big(\mu^2\,\hat{\sigma}^{i}_{ps}\hat{\sigma}^{j}_{sp} + \nu^2\,\hat{\sigma}^{i}_{ps'}\hat{\sigma}^{j}_{s'p} \\
&\qquad + \alpha^2\,\hat{\sigma}^{i}_{p's}\hat{\sigma}^{j}_{sp'} + \beta^2\,\hat{\sigma}^{i}_{p's'}\hat{\sigma}^{j}_{s'p'}\Big)\frac{1}{R_{ij}^3}.
\end{split}
\label{eq:H3}
\end{equation}
In each expression, the operator $\hat{\sigma}^i_{ab}$ takes the $i^{th}$ atom from state $a$ to state $b$, $R_{ij}$ is the distance between the $i^{th}$ and $j^{th}$ atoms, and $\mu$, $\nu$, $\alpha$, and $\beta$ are the dipole moments coupling ${p\leftrightarrow s}$, ${p\leftrightarrow s'}$, ${p'\leftrightarrow s}$, and ${p'\leftrightarrow s'}$, respectively.
In both cases, the first sum is the field-tuned exchange and the remaining sums are the hopping terms. The three-body matrix element in Eq.~(\ref{eq:H3}) is
\begin{equation}
V_3^{ijk} = \frac{\mu\nu(\beta\nu + \alpha\mu)}{\delta\,R_{jk}^3\,R_{ik}^3},
\label{eq:V3}
\end{equation}
which results from summing over all possible paths to the final state~\cite{spielman_QuantumManybodyScars_2024}. 

The two-body field-tuned and hopping interactions both scale with $R^{-3}$, while the three-body field-tuned interaction effectively scales as $R^{-6}$. Spielman \textit{et al.} showed that the combination of strong hopping interactions and a weak field-tuned interaction for the three-body case gives rise to quantum many-body scar states~\cite{spielman_QuantumManybodyScars_2024}. 

In our model, we place $N$ atoms in a linear array, with initial state $\ket{pp\cdots p}$. We choose our linear array spacing so that the field-tuned two- and three-body interactions have matrix elements with similar magnitude. As is evident from Eqs.~(\ref{eq:H2}) and (\ref{eq:H3}), we also ignore the angular dependence of the dipole-dipole interaction.

\section{Results}\label{sec:results}

\begin{figure}
\centering
\includegraphics[width=\columnwidth]{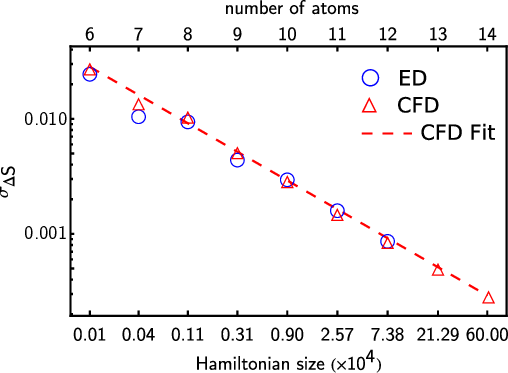}
\caption{Dependence of $\sigma_{\Delta S}$ on the of Hilbert-space dimension $D$ for the two-body  \spps{} model. Blue circles are exact-diagonalization (ED) results for the available small-system sizes; red triangles are obtained with the CFD sampling method. The dashed lines is a power-law fit to the sampled results, consistent with a $D^{-1/2}$ dependence.}
\label{fig:deviationtwo}
\end{figure}

\begin{figure}
\centering
\includegraphics[width=\columnwidth]{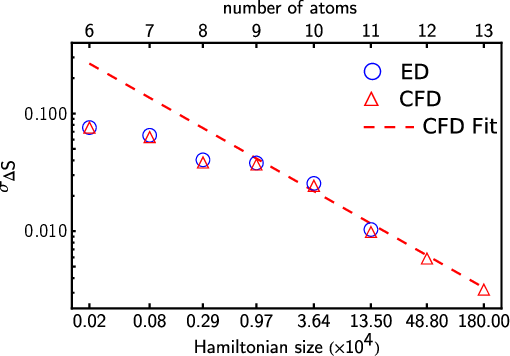}
\caption{Dependence of $\sigma_{\Delta S}$ on the of Hilbert-space dimension $D$ for the three-body  \spps{} model. Blue circles are exact-diagonalization (ED) results for the available small-system sizes; red triangles are obtained with the CFD sampling method. The dashed lines is a power-law fit to the sampled results for nine atoms and greater, consistent with a $D^{-1/2}$ dependence.}
\label{fig:deviationthree}
\end{figure}

Figures~\ref{fig:deviationtwo} and~\ref{fig:deviationthree} report $\sigma_{\Delta S}$ as a function of the Hilbert-space dimension $D$ for the two- and three-body regimes, respectively. In the two-body regime, ED is carried out up to $D=7.38\!\times\!10^{4}$ (12 atoms); the present method extends the calculation to larger Hilbert-space dimensions reaching $D=6.13\!\times\!10^{5}$ (14 atoms). The ED and CFD data are in good agreement. A fit to the CFD sampling results shows a decay of $\sigma_{\Delta S}$ consistent with the $D^{-1/2}$ scaling as reported in previous studies~\cite{beugeling_FinitesizeScalingEigenstate_2014, ikeda_FinitesizeScalingAnalysis_2013}.

In the three-body regime, ED is carried out up to $D=1.35\!\times\!10^{5}$ (11 atoms), while the CFD pipeline extends the calculation to $D=1.8\!\times\!10^{6}$ (13 atoms). We fit the CFD sampling results only for greater than nine atoms. At small numbers of atoms, the three-body interaction suffers greatly from finite size effects and the behavior has not yet converged, as discussed in Ref.~\cite{spielman_QuantumManybodyScars_2024}. The fit is consistent with $D^{-1/2}$ scaling of $\sigma_{\Delta S}$, as in the two-body case. However, it is notable here that the ED results by themselves are inconclusive.

\section{Discussion and conclusion}\label{sec:conclusion}

We have introduced a scalable test of weak ETH and applied it to the \spps{} Rydberg model at Hilbert-space dimensions extending well beyond the largest sizes accessible to ED. The decay of $\sigma_{\Delta S}$ was found to scale as $D^{-1/2}$, consistent with previous numerical work~\cite{beugeling_FinitesizeScalingEigenstate_2014, ikeda_FinitesizeScalingAnalysis_2013,rigol_BreakdownThermalizationFinite_2009}. Huang, however, has shown analytically that $\sigma_{\Delta S}$ should scale as the reciprocal of system size, in contrast to our result~\cite{huang_deviation_2024}. Huang speculates that numerical results to date cannot access large enough system sizes to see this behavior; we plan to test our method at even larger system sizes to search for any deviation from the observed $D^{-1/2}$ scaling.

One way to test the predictions of ETH is to calculate the expected thermal state and compare to experimental measurement, as has recently been done in~\cite{zhao_ObservationQuantumThermalization_2025} and~\cite{spielman_ThermalizationStarkManifold_2026}. Dynamical quantum typicality is an efficient way to calculate the thermal state, particularly for large, sparse Hamiltonians, however it relies on the system obeying at least wETH~\cite{reimann_TypicalityGeneralizedMicrocanonical_2007, bartsch_DynamicalTypicalityQuantum_2009,steinigeweg_PushingLimitsEigenstate_2014,heitmann_SelectedApplicationsTypicality_2020, reimann_DynamicalTypicalityIsolated_2018,mitric_DynamicalQuantumTypicality_2025}. Thus, our present method of testing wETH can validate such thermal state calculations.
 
More broadly, we expect that random sampling of the eigenstates should be useful whenever statistical information about the eigenstates is of value. For example, recent work on subtle correlations among mid-spectrum eigenstates might benefit.

\begin{acknowledgments}
This work was supported by the National Science Foundation under Grants No.~2011583, No.~2011610,  and No.~2427091. S.~E.~Spielman is supported by the National Science Foundation Graduate Research Fellowship under Grant No.~2334429. This work used the Delta system at the National Center for Supercomputing Applications through allocation PHY230142 from the Advanced Cyberinfrastructure Coordination Ecosystem: Services \& Support (ACCESS) program, supported by National Science Foundation grants \#2138259, \#2138286, \#2138307, \#2137603, and \#2138296.
\end{acknowledgments}

\appendix
\label{app:app}
\section{Implementation Details}\label{app:impl}

\emph{Moments and probes.}---Since the KPM density only guides where in the spectrum target energies are drawn and how wide each bracket is, it does not need to resolve fine spectral structure; the accuracy of the extracted eigenpairs depends entirely on the subsequent CFD solve. We find $M=100$ Chebyshev moments and $R=1$ Gaussian random vector sufficient for a smooth, well-normalized $\rho(E)$ at every Hamiltonian size, which makes the KPM pre-pass cost a small fraction of a single
CFD solve.

\emph{Adaptive bracket.}---The choice of the bracket size is the
most important decision in this method. For a bracket containing $N_T$ target eigenvalues, CFD requires an oversampled search space, and Ref.~\cite{pieper_HighperformanceImplementationChebyshev_2016} recommends $2\le N_S/N_T\le 4$. Wide brackets increase the number of search vectors, each of length $D$, but require a lower polynomial degree $N_P$. Narrow brackets require fewer search
vectors but a sharper, higher-degree filter to isolate the eigenpairs of $\widetilde H$. The memory cost is dominated by the $N_S$ search vectors; at $D\approx1.8\times10^6$, storing $N_S=2000$ search vectors in double precision requires $28.8$~GB of device memory. Single precision would halve this to $14.4$~GB and could extend the method to larger $D$, but we use double precision on GPUs with sufficient memory because 
at large $D$ the spacing between neighboring eigenvalues of
$\widetilde H$ approaches single-precision resolution, and the Ritz pairs then fail to converge. We therefore set $N_S=2N_T$, the smallest oversampling in the recommended range, which we found sufficient for robust convergence across all system sizes studied while remaining within device memory.

\emph{Polynomial degree.}---Rather than computing $N_P$ analytically, we determined it empirically for each Hilbert-space dimension by testing convergence of the Ritz pairs against the residual tolerance; the resulting values are reported in Table~\ref{tab:params}.

\emph{Residual control.}---We accept a Ritz pair $(\theta_j,\mathbf v_j)$, with $\mathbf v_j$ normalized, if
$\|H\mathbf v_j-\theta_j\mathbf v_j\|_2\le\tau_{\rm res}|\theta_j|$ with $\tau_{\rm res}=10^{-3}$.  For representative system sizes in both regimes, tightening $\tau_{\rm res}$ to $10^{-5}$ produced little to no change in
$\Delta(\hat S)$, confirming that $10^{-3}$ is sufficient for the wETH diagnostic. To avoid retaining duplicate eigenpairs, accepted Ritz pairs from the same Rayleigh--Ritz step whose eigenvalues differ by less than $10^{-4}$ in relative terms are treated as duplicates, and only the pair with the smaller residual norm is kept.

\emph{Iteration stopping criterion.}---Although $N_T$ sets the target number of eigenvalues in the bracket, we do not force each CFD calculation to converge exactly $N_T$ eigenpairs. Instead, after constructing the search subspace, we run at most a fixed number of CFD iterations and retain only the Ritz pairs that satisfy the residual criterion, stopping early once $n_\ell$ pairs have been accepted. For Hilbert-space dimensions $D>2\times10^5$ we allow five iterations per bracket; for smaller systems, where the cost is lower, we allow eight. This may not be universal. Empirically, the largest improvement in the number and quality of accepted Ritz pairs occurs during the first few iterations, while the convergence rate slows substantially after roughly five iterations for the larger systems. At the largest $D$, fewer than $n_\ell$ pairs
typically converge within the iteration limit, and we keep all of them,
so $n_\ell$ acts as an upper limit. This stopping rule balances
eigenpair accuracy with the need to accumulate many independent samples
for the wETH statistic.

\emph{Sampling scheme.}---For each Hamiltonian and regime,
the reported statistics are built from $2\times10^3$ sampled eigenstate
expectation values. The production sampling was distributed across
eight NVIDIA H200 GPUs by loading the same Hamiltonian on each GPU and
drawing brackets independently on each device. This parallelization
does not split the Hamiltonian or change the CFD algorithm; it simply
makes sampling more efficient, because the brackets are independent and
the retained expectation values are combined only at the final
statistics stage.

\emph{GPU implementation with Ginkgo.}---All operations on vectors of
length $D$ use the Ginkgo library~\cite{anzt_GinkgoModernLinear_2022}. The Hamiltonian is
copied once to the GPU and stays there, together with the search vectors. Since sparse matrix--vector products dominate the
cost of both the KPM and CFD, running them
on the GPU speeds up the most expensive part of the method. The small
dense eigenvalue problems are solved on the host with LAPACK.
 
\emph{Spectral bounds.}---We obtain $E_{\min}$ and $E_{\max}$ with the
$k$-step Lanczos method~\cite{zhou_BoundingSpectrumLarge_2011}. Starting from a
normalized Gaussian random vector, $k=30$ Lanczos steps build a
$k\times k$ tridiagonal matrix, whose extremal eigenvalues, widened by
the Lanczos residual estimates, give the bounds at the cost of only $k$
matrix--vector products. 

\section{CFD parameters by system size}\label{sec:params}
 
Table~\ref{tab:params} lists the Hilbert-space dimension $D$ and the converged adaptive-window CFD parameters $N_T$ (target eigenvalues in the bracket), $N_S$ (search vectors), and $N_P$ (polynomial filter order) for each atom number and regime studied. The table illustrates how the adaptive logic keeps $N_S$ and $N_P$ within
hardware budget as $D$ grows.
\onecolumngrid
\begin{table}[H]
\setlength{\columnwidth}{\textwidth}%
\setlength{\linewidth}{\textwidth}%
\setlength{\hsize}{\textwidth}%
\caption{CFD parameters for each system size. $N$ is the number of atoms; $D$ is the Hilbert-space dimension; $N_T$, $N_S$, and $N_P$ are the target eigenvalue count, number of search vectors, and polynomial filter order. Results are shown for the two-body and three-body regimes. ED entries were computed by full exact diagonalization; CFD entries uses the method given in Section~\ref{sec:method}.}
\label{tab:params}
\begin{ruledtabular}
\begin{tabular}{crrrrrcrrrrr}
 & \multicolumn{5}{c}{Two-body regime} &
 & \multicolumn{5}{c}{Three-body regime} \\
\noalign{\smallskip}
$N$ & $D$ & $N_T$ & $N_S$ & $N_P$ & Method &
    & $D$ & $N_T$ & $N_S$ & $N_P$ & Method \\
\hline
6& 141 &50&100&100&ED && 211&50&100&100&ED\\
 7&       393 & 150 & 300 &  100 & ED  & &       841 & 150 & 300 &  100 & ED  \\
 8 &      1,107 & 300 & 600 &  100 & ED  & &      2,857 & 300 & 600 &  300 & ED  \\
 9 &      3,139 & 300 & 600 &  100 & ED  & &      9,745 & 300 & 600 &  400 & ED  \\
10 &      8,953 & 300 & 600 &  400 & ED  & &     36,421 & 500 & 1000 & 700 & ED  \\
11 &     25,653 & 500 & 1000 & 700 & ED  & &    134,971 & 1500 & 3000 & 1000 & ED  \\
12 &     73,789 & 1500 & 3000 & 1000 & ED  & &    540,000 & 1000 & 4000 & 4500 & CFD \\
13 &    212,941 & 1500 & 3000 & 3500 & CFD & &   1,800,000 & 450 & 900 & 27000 & CFD \\
14 &    616,227 & 1000 &  2000 &  4500 & CFD & &       6,700,000 & 100 & 200 & 37000 & CFD \\
15 &   1,787,607 &  450 &  900 &  27000 & CFD & & LIMIT & $\otimes$ & $\otimes$ & $\otimes$ & $\otimes$ \\
16 & 5,196,627 & 100 & 200 & 37000 & CFD &&\\
17 & LIMIT & $\otimes$ & $\otimes$ & $\otimes$ & $\otimes$ \\
\end{tabular}
\end{ruledtabular}
\end{table}
\twocolumngrid

\bibliography{wETH-stat}

@article{abanin_ColloquiumManybodyLocalization_2019,
  title = {Colloquium: {{Many}}-Body Localization, Thermalization, and Entanglement},
  shorttitle = {Colloquium},
  author = {Abanin, Dmitry A. and Altman, Ehud and Bloch, Immanuel and Serbyn, Maksym},
  year = 2019,
  month = may,
  journal = {Rev. Mod. Phys.},
  volume = {91},
  number = {2},
  pages = {021001},
  doi = {10.1103/RevModPhys.91.021001},
  urldate = {2019-11-17}
}

@article{anzt_GinkgoModernLinear_2022,
  title = {Ginkgo: {{A Modern Linear Operator Algebra Framework}} for {{High Performance Computing}}},
  shorttitle = {Ginkgo},
  author = {Anzt, Hartwig and Cojean, Terry and Flegar, Goran and G{\"o}bel, Fritz and Gr{\"u}tzmacher, Thomas and Nayak, Pratik and Ribizel, Tobias and Tsai, Yuhsiang Mike and {Quintana-Ort{\'i}}, Enrique S.},
  year = 2022,
  month = feb,
  journal = {ACM Transactions on Mathematical Software},
  volume = {48},
  number = {1},
  pages = {2:1--2:33},
  issn = {0098-3500},
  doi = {10.1145/3480935},
  urldate = {2022-02-17},
  copyright = {All rights reserved}
}

@article{bartsch_DynamicalTypicalityQuantum_2009,
  title = {Dynamical {{Typicality}} of {{Quantum Expectation Values}}},
  author = {Bartsch, Christian and Gemmer, Jochen},
  year = 2009,
  month = mar,
  journal = {Phys. Rev. Lett.},
  volume = {102},
  number = {11},
  pages = {110403},
  publisher = {American Physical Society},
  doi = {10.1103/PhysRevLett.102.110403},
  urldate = {2024-09-09}
}

@article{beugeling_FinitesizeScalingEigenstate_2014,
  title = {Finite-Size Scaling of Eigenstate Thermalization},
  author = {Beugeling, W. and Moessner, R. and Haque, Masudul},
  year = 2014,
  month = apr,
  journal = {Phys. Rev. E},
  volume = {89},
  number = {4},
  pages = {042112},
  publisher = {American Physical Society},
  doi = {10.1103/PhysRevE.89.042112},
  urldate = {2025-06-29}
}

@article{deutsch_QuantumStatisticalMechanics_1991,
  title = {Quantum Statistical Mechanics in a Closed System},
  author = {Deutsch, J. M.},
  year = 1991,
  month = feb,
  journal = {Phys. Rev. A},
  volume = {43},
  number = {4},
  pages = {2046--2049},
  publisher = {American Physical Society},
  doi = {10.1103/PhysRevA.43.2046},
  urldate = {2021-07-12}
}

@article{dooley_ExtremeManybodyScarring_2022,
  title = {Extreme Many-Body Scarring in a Quantum Spin Chain via Weak Dynamical Constraints},
  author = {Dooley, Shane and Kells, Graham},
  year = 2022,
  month = apr,
  journal = {Phys. Rev. B},
  volume = {105},
  number = {15},
  pages = {155127},
  publisher = {American Physical Society},
  doi = {10.1103/PhysRevB.105.155127},
  urldate = {2024-09-17}
}

@article{faoro_BorromeanThreebodyFRET_2015,
  title = {Borromean Three-Body {{FRET}} in Frozen {{Rydberg}} Gases},
  author = {Faoro, R. and Pelle, B. and Zuliani, A. and Cheinet, P. and Arimondo, E. and Pillet, P.},
  year = 2015,
  month = sep,
  journal = {Nature Communications},
  volume = {6},
  number = {1},
  pages = {1--7},
  issn = {2041-1723},
  doi = {10.1038/ncomms9173},
  urldate = {2020-02-21},
  copyright = {2015 The Author(s)},
  langid = {english}
}

@article{hahn_EigenstateCorrelationsEigenstate_2024,
  title = {Eigenstate {{Correlations}}, the {{Eigenstate Thermalization Hypothesis}}, and {{Quantum Information Dynamics}} in {{Chaotic Many-Body Quantum Systems}}},
  author = {Hahn, Dominik and Luitz, David J. and Chalker, J. T.},
  year = 2024,
  month = aug,
  journal = {Phys. Rev. X},
  volume = {14},
  number = {3},
  pages = {031029},
  publisher = {American Physical Society},
  doi = {10.1103/PhysRevX.14.031029},
  urldate = {2026-06-22}
}

@article{haque_EntanglementMidspectrumEigenstates_2022,
  title = {Entanglement of Midspectrum Eigenstates of Chaotic Many-Body Systems: {{Reasons}} for Deviation from Random Ensembles},
  shorttitle = {Entanglement of Midspectrum Eigenstates of Chaotic Many-Body Systems},
  author = {Haque, Masudul and McClarty, Paul A. and Khaymovich, Ivan M.},
  year = 2022,
  month = jan,
  journal = {Phys. Rev. E},
  volume = {105},
  number = {1},
  pages = {014109},
  publisher = {American Physical Society},
  doi = {10.1103/PhysRevE.105.014109},
  urldate = {2026-06-22}
}

@article{heitmann_SelectedApplicationsTypicality_2020,
  title = {Selected Applications of Typicality to Real-Time Dynamics of Quantum Many-Body Systems},
  author = {Heitmann, Tjark and Richter, Jonas and Schubert, Dennis and Steinigeweg, Robin},
  year = 2020,
  month = may,
  journal = {Zeitschrift f\"ur Naturforschung A},
  volume = {75},
  number = {5},
  pages = {421--432},
  publisher = {De Gruyter},
  issn = {1865-7109},
  doi = {10.1515/zna-2020-0010},
  urldate = {2024-09-19},
  langid = {english}
}

@article{huang_deviation_2024,
  title = {Deviation from Maximal Entanglement for Mid-Spectrum Eigenstates of Local {{Hamiltonians}}},
  author = {Huang, Yichen},
  year = 2024,
  journal = {IEEE J. Sel. Areas Inf. Theory},
  volume = {5},
  eprint = {2202.01173},
  primaryclass = {quant-ph},
  pages = {694--701},
  issn = {2641-8770},
  doi = {10.1109/JSAIT.2024.3487856},
  urldate = {2026-09-25},
  archiveprefix = {arXiv}
}

@article{huang_FinitesizeScalingAnalysis_2022,
  title = {Finite-Size Scaling Analysis of Eigenstate Thermalization},
  author = {Huang, Yichen},
  year = 2022,
  month = mar,
  journal = {Annals of Physics},
  volume = {438},
  pages = {168761},
  issn = {0003-4916},
  doi = {10.1016/j.aop.2022.168761},
  urldate = {2026-06-23}
}

@article{ikeda_FinitesizeScalingAnalysis_2013,
  title = {Finite-Size Scaling Analysis of the Eigenstate Thermalization Hypothesis in a One-Dimensional Interacting {{Bose}} Gas},
  author = {Ikeda, Tatsuhiko N. and Watanabe, Yu and Ueda, Masahito},
  year = 2013,
  month = jan,
  journal = {Phys. Rev. E},
  volume = {87},
  number = {1},
  pages = {012125},
  publisher = {American Physical Society},
  doi = {10.1103/PhysRevE.87.012125},
  urldate = {2025-06-29}
}

@article{ikeda_HowAccuratelyCan_2015,
  title = {How Accurately Can the Microcanonical Ensemble Describe Small Isolated Quantum Systems?},
  author = {Ikeda, Tatsuhiko N. and Ueda, Masahito},
  year = 2015,
  month = aug,
  journal = {Phys. Rev. E},
  volume = {92},
  number = {2},
  pages = {020102},
  publisher = {American Physical Society},
  doi = {10.1103/PhysRevE.92.020102},
  urldate = {2026-09-28}
}

@article{kim_TestingWhetherAll_2014,
  title = {Testing Whether All Eigenstates Obey the Eigenstate Thermalization Hypothesis},
  author = {Kim, Hyungwon and Ikeda, Tatsuhiko N. and Huse, David A.},
  year = 2014,
  month = nov,
  journal = {Phys. Rev. E},
  volume = {90},
  number = {5},
  pages = {052105},
  publisher = {American Physical Society},
  doi = {10.1103/physreve.90.052105},
  urldate = {2022-02-07}
}

@article{kliczkowski_AverageEntanglementEntropy_2023,
  title = {Average Entanglement Entropy of Midspectrum Eigenstates of Quantum-Chaotic Interacting {{Hamiltonians}}},
  author = {Kliczkowski, M. and {\'S}wietek, R. and Vidmar, L. and Rigol, M.},
  year = 2023,
  month = jun,
  journal = {Phys. Rev. E},
  volume = {107},
  number = {6},
  pages = {064119},
  publisher = {American Physical Society},
  doi = {10.1103/PhysRevE.107.064119},
  urldate = {2026-06-23}
}

@article{langlett_EntanglementPatternsQuantum_2025,
  title = {Entanglement {{Patterns}} of {{Quantum Chaotic Hamiltonians}} with a {{Scalar U}}(1) {{Charge}}},
  author = {Langlett, Christopher M. and {Rodriguez-Nieva}, Joaquin F.},
  year = 2025,
  month = jun,
  journal = {Phys. Rev. Lett.},
  volume = {134},
  number = {23},
  pages = {230402},
  publisher = {American Physical Society},
  doi = {10.1103/xxlq-d1sw},
  urldate = {2026-06-23}
}

@article{liu_TimeDependenceFewBody_2020,
  title = {Time {{Dependence}} of {{Few-Body F}}\textbackslash "orster {{Interactions}} among {{Ultracold Rydberg Atoms}}},
  author = {Liu, Zhimin Cheryl and Inman, Nina P. and Carroll, Thomas J. and Noel, Michael W.},
  year = 2020,
  month = mar,
  journal = {Phys. Rev. Lett.},
  volume = {124},
  number = {13},
  pages = {133402},
  publisher = {American Physical Society},
  doi = {10.1103/PhysRevLett.124.133402},
  urldate = {2025-12-20}
}

@article{luschen_ObservationSlowDynamics_2017,
  title = {Observation of {{Slow Dynamics}} near the {{Many-Body Localization Transition}} in {{One-Dimensional Quasiperiodic Systems}}},
  author = {L{\"u}schen, Henrik P. and Bordia, Pranjal and Scherg, Sebastian and Alet, Fabien and Altman, Ehud and Schneider, Ulrich and Bloch, Immanuel},
  year = 2017,
  month = dec,
  journal = {Phys. Rev. Lett.},
  volume = {119},
  number = {26},
  pages = {260401},
  issn = {0031-9007, 1079-7114},
  doi = {10.1103/PhysRevLett.119.260401},
  urldate = {2024-10-28},
  copyright = {https://link.aps.org/licenses/aps-default-license},
  langid = {english}
}

@article{mitric_DynamicalQuantumTypicality_2025,
  title = {Dynamical Quantum Typicality: {{Simple}} Method for Investigating Transport Properties Applied to the {{Holstein}} Model},
  shorttitle = {Dynamical Quantum Typicality},
  author = {Mitri{\'c}, Petar},
  year = 2025,
  month = may,
  journal = {Phys. Rev. B},
  volume = {111},
  number = {19},
  pages = {195140},
  publisher = {American Physical Society},
  doi = {10.1103/PhysRevB.111.195140},
  urldate = {2025-06-13}
}

@article{nandkishore_ManyBodyLocalizationThermalization_2015,
  title = {Many-{{Body Localization}} and {{Thermalization}} in {{Quantum Statistical Mechanics}}},
  author = {Nandkishore, Rahul and Huse, David A.},
  year = 2015,
  journal = {Annual Review of Condensed Matter Physics},
  volume = {6},
  number = {1},
  pages = {15--38},
  doi = {10.1146/annurev-conmatphys-031214-014726},
  urldate = {2016-05-10}
}

@article{pieper_HighperformanceImplementationChebyshev_2016,
  title = {High-Performance Implementation of {{Chebyshev}} Filter Diagonalization for Interior Eigenvalue Computations},
  author = {Pieper, Andreas and Kreutzer, Moritz and Alvermann, Andreas and Galgon, Martin and Fehske, Holger and Hager, Georg and Lang, Bruno and Wellein, Gerhard},
  year = 2016,
  month = nov,
  journal = {Journal of Computational Physics},
  volume = {325},
  pages = {226--243},
  issn = {00219991},
  doi = {10.1016/j.jcp.2016.08.027},
  urldate = {2025-08-18},
  langid = {english}
}

@article{reimann_DynamicalTypicalityApproach_2018,
  title = {Dynamical {{Typicality Approach}} to {{Eigenstate Thermalization}}},
  author = {Reimann, Peter},
  year = 2018,
  month = jun,
  journal = {Phys. Rev. Lett.},
  volume = {120},
  number = {23},
  pages = {230601},
  publisher = {American Physical Society},
  doi = {10.1103/PhysRevLett.120.230601},
  urldate = {2024-09-17}
}

@article{reimann_DynamicalTypicalityIsolated_2018,
  title = {Dynamical Typicality of Isolated Many-Body Quantum Systems},
  author = {Reimann, Peter},
  year = 2018,
  month = jun,
  journal = {Phys. Rev. E},
  volume = {97},
  number = {6},
  eprint = {1805.07085},
  primaryclass = {cond-mat},
  pages = {062129},
  issn = {2470-0045, 2470-0053},
  doi = {10.1103/PhysRevE.97.062129},
  urldate = {2025-05-29},
  archiveprefix = {arXiv}
}

@article{reimann_TypicalityGeneralizedMicrocanonical_2007,
  title = {Typicality for {{Generalized Microcanonical Ensembles}}},
  author = {Reimann, Peter},
  year = 2007,
  month = oct,
  journal = {Phys. Rev. Lett.},
  volume = {99},
  number = {16},
  pages = {160404},
  publisher = {American Physical Society},
  doi = {10.1103/PhysRevLett.99.160404},
  urldate = {2024-11-04}
}

@article{rigol_BreakdownThermalizationFinite_2009,
  title = {Breakdown of {{Thermalization}} in {{Finite One-Dimensional Systems}}},
  author = {Rigol, Marcos},
  year = 2009,
  month = sep,
  journal = {Phys. Rev. Lett.},
  volume = {103},
  number = {10},
  pages = {100403},
  publisher = {American Physical Society},
  doi = {10.1103/PhysRevLett.103.100403},
  urldate = {2025-06-29}
}

@article{rigol_ThermalizationItsMechanism_2008,
  title = {Thermalization and Its Mechanism for Generic Isolated Quantum Systems},
  author = {Rigol, Marcos and Dunjko, Vanja and Olshanii, Maxim},
  year = 2008,
  month = apr,
  journal = {Nature},
  volume = {452},
  number = {7189},
  pages = {854--858},
  publisher = {Nature Publishing Group},
  issn = {1476-4687},
  doi = {10.1038/nature06838},
  urldate = {2021-03-24},
  langid = {english}
}

@article{rodriguez-nieva_QuantifyingQuantumChaos_2024,
  title = {Quantifying {{Quantum Chaos}} through {{Microcanonical Distributions}} of {{Entanglement}}},
  author = {{Rodriguez-Nieva}, Joaquin F. and Jonay, Cheryne and Khemani, Vedika},
  year = 2024,
  month = jul,
  journal = {Phys. Rev. X},
  volume = {14},
  number = {3},
  pages = {031014},
  publisher = {American Physical Society},
  doi = {10.1103/PhysRevX.14.031014},
  urldate = {2026-06-23}
}

@article{sala_ErgodicityBreakingArising_2020,
  title = {Ergodicity {{Breaking Arising}} from {{Hilbert Space Fragmentation}} in {{Dipole-Conserving Hamiltonians}}},
  author = {Sala, Pablo and Rakovszky, Tibor and Verresen, Ruben and Knap, Michael and Pollmann, Frank},
  year = 2020,
  month = feb,
  journal = {Phys. Rev. X},
  volume = {10},
  number = {1},
  pages = {011047},
  publisher = {American Physical Society},
  doi = {10.1103/PhysRevX.10.011047},
  urldate = {2022-03-22}
}

@article{spielman_QuantumManybodyScars_2024,
  title = {Quantum Many-Body Scars in Few-Body Dipole-Dipole Interactions},
  author = {Spielman, Sarah E. and Handian, Alicia and Inman, Nina P. and Carroll, Thomas J. and Noel, Michael W.},
  year = 2024,
  month = nov,
  journal = {Phys. Rev. Res.},
  volume = {6},
  number = {4},
  pages = {043086},
  publisher = {American Physical Society},
  doi = {10.1103/PhysRevResearch.6.043086},
  urldate = {2026-07-07}
}

@article{spielman_ThermalizationStarkManifold_2026,
  title = {Thermalization within a {{Stark}} Manifold through {{Rydberg}} Atom Interactions},
  author = {Spielman, Sarah E. and Thomas, Sage M. and Teofilovska, Maja and van Blerkom, Annick C. and {Bauroth-Sherman}, Juniper J. and Chlanda, Nicolaus A. and Conley, Hannah S. and Conte, Philip A. and Kirk, Aidan D. and Carroll, Thomas J. and Noel, Michael W.},
  year = 2026,
  month = jul,
  journal = {arXiv},
  eprint = {2512.22110},
  primaryclass = {quant-ph},
  publisher = {arXiv},
  doi = {10.48550/arXiv.2512.22110},
  urldate = {2026-07-21},
  archiveprefix = {arXiv}
}

@article{srednicki_chaos_1994,
  title = {Chaos and Quantum Thermalization},
  author = {Srednicki, Mark},
  year = 1994,
  month = aug,
  journal = {Phys. Rev. E},
  volume = {50},
  number = {2},
  pages = {888--901},
  publisher = {American Physical Society},
  doi = {10.1103/PhysRevE.50.888},
  urldate = {2026-09-17}
}

@article{stathopoulos_BlockOrthogonalizationProcedure_2002,
  title = {A {{Block Orthogonalization Procedure}} with {{Constant Synchronization Requirements}}},
  author = {Stathopoulos, Andreas and Wu, Kesheng},
  year = 2002,
  month = jan,
  journal = {SIAM J. Sci. Comput.},
  volume = {23},
  number = {6},
  pages = {2165--2182},
  publisher = {{Society for Industrial and Applied Mathematics}},
  issn = {1064-8275},
  doi = {10.1137/S1064827500370883},
  urldate = {2025-10-24}
}

@article{steinigeweg_PushingLimitsEigenstate_2014,
  title = {Pushing the {{Limits}} of the {{Eigenstate Thermalization Hypothesis}} towards {{Mesoscopic Quantum Systems}}},
  author = {Steinigeweg, R. and Khodja, A. and Niemeyer, H. and Gogolin, C. and Gemmer, J.},
  year = 2014,
  month = apr,
  journal = {Phys. Rev. Lett.},
  volume = {112},
  number = {13},
  pages = {130403},
  publisher = {American Physical Society},
  doi = {10.1103/PhysRevLett.112.130403},
  urldate = {2024-09-16}
}

@article{sugimoto_EigenstateThermalizationLongRange_2022,
  title = {Eigenstate {{Thermalization}} in {{Long-Range Interacting Systems}}},
  author = {Sugimoto, Shoki and Hamazaki, Ryusuke and Ueda, Masahito},
  year = 2022,
  month = jul,
  journal = {Phys. Rev. Lett.},
  volume = {129},
  number = {3},
  pages = {030602},
  publisher = {American Physical Society},
  doi = {10.1103/PhysRevLett.129.030602},
  urldate = {2024-05-24}
}

@article{sugimoto_TestEigenstateThermalization_2021,
  title = {Test of the {{Eigenstate Thermalization Hypothesis Based}} on {{Local Random Matrix Theory}}},
  author = {Sugimoto, Shoki and Hamazaki, Ryusuke and Ueda, Masahito},
  year = 2021,
  month = mar,
  journal = {Phys. Rev. Lett.},
  volume = {126},
  number = {12},
  pages = {120602},
  publisher = {American Physical Society},
  doi = {10.1103/PhysRevLett.126.120602},
  urldate = {2025-09-02}
}

@article{tretyakov_ObservationBorromeanThreeBody_2017,
  title = {Observation of the {{Borromean Three-Body F\"orster Resonances}} for {{Three Interacting Rb Rydberg Atoms}}},
  author = {Tretyakov, D. B. and Beterov, I. I. and Yakshina, E. A. and Entin, V. M. and Ryabtsev, I. I. and Cheinet, P. and Pillet, P.},
  year = 2017,
  month = oct,
  journal = {Phys. Rev. Lett.},
  volume = {119},
  number = {17},
  pages = {173402},
  doi = {10.1103/PhysRevLett.119.173402},
  urldate = {2018-04-08}
}

@article{turner_QuantumScarredEigenstates_2018,
  title = {Quantum Scarred Eigenstates in a {{Rydberg}} Atom Chain: {{Entanglement}}, Breakdown of Thermalization, and Stability to Perturbations},
  shorttitle = {Quantum Scarred Eigenstates in a {{Rydberg}} Atom Chain},
  author = {Turner, C. J. and Michailidis, A. A. and Abanin, D. A. and Serbyn, M. and Papi{\'c}, Z.},
  year = 2018,
  month = oct,
  journal = {Phys. Rev. B},
  volume = {98},
  number = {15},
  pages = {155134},
  doi = {10.1103/PhysRevB.98.155134},
  urldate = {2018-10-24}
}

@article{turner_WeakErgodicityBreaking_2018,
  title = {Weak Ergodicity Breaking from Quantum Many-Body Scars},
  author = {Turner, C. J. and Michailidis, A. A. and Abanin, D. A. and Serbyn, M. and Papi{\'c}, Z.},
  year = 2018,
  month = jul,
  journal = {Nature Phys},
  volume = {14},
  number = {7},
  pages = {745--749},
  publisher = {Nature Publishing Group},
  issn = {1745-2481},
  doi = {10.1038/s41567-018-0137-5},
  urldate = {2022-02-07},
  copyright = {2018 The Author(s)},
  langid = {english}
}

@article{weisse_KernelPolynomialMethod_2006,
  title = {The Kernel Polynomial Method},
  author = {Wei{\ss}e, Alexander and Wellein, Gerhard and Alvermann, Andreas and Fehske, Holger},
  year = 2006,
  month = mar,
  journal = {Rev. Mod. Phys.},
  volume = {78},
  number = {1},
  pages = {275--306},
  issn = {0034-6861, 1539-0756},
  doi = {10.1103/RevModPhys.78.275},
  urldate = {2025-06-17},
  copyright = {http://link.aps.org/licenses/aps-default-license},
  langid = {english}
}

@article{zhao_ObservationQuantumThermalization_2025,
  title = {Observation of {{Quantum Thermalization Restricted}} to {{Hilbert Space Fragments}} and {{Z}} 2 k {{Scars}}},
  author = {Zhao, Luheng and Datla, Prithvi Raj and Tian, Weikun and Aliyu, Mohammad Mujahid and Loh, Huanqian},
  year = 2025,
  month = feb,
  journal = {Phys. Rev. X},
  volume = {15},
  number = {1},
  pages = {011035},
  issn = {2160-3308},
  doi = {10.1103/PhysRevX.15.011035},
  urldate = {2025-05-30},
  langid = {english}
}

@article{zhou_BoundingSpectrumLarge_2011,
  title = {Bounding the Spectrum of Large {{Hermitian}} Matrices},
  author = {Zhou, Yunkai and Li, Ren-Cang},
  year = 2011,
  month = aug,
  journal = {Linear Algebra and its Applications},
  series = {Special {{Issue}}: {{Dedication}} to {{Pete Stewart}} on the Occasion of His 70th Birthday},
  volume = {435},
  number = {3},
  pages = {480--493},
  issn = {0024-3795},
  doi = {10.1016/j.laa.2010.06.034},
  urldate = {2025-10-08}
}

\end{document}